\documentstyle[pra,aps,psfig,multicol]{revtex}
\draft 
\begin{document}
\title{Non-holonomic Quantum Devices} 
\author{ V.~M. Akulin,$^1$ V.~Gershkovich,$^2$ and G.~Harel$^3$} 
\address{ $^1$Laboratoire Aim\'e Cotton, CNRS II, 
B\^atiment 505, 91405 Orsay Cedex, France\\
$^2$Institut des Hautes Etudes Scientifiques, Bures-sur-Yvette, France\\
$^3$Department of Physics and Astronomy, Vrije Universiteit, 
De Boelelaan 1081, 1081 HV Amsterdam, The Netherlands}
\date{December 19, 2000}
\maketitle
\widetext
\begin{abstract}
We analyze the possibility and efficiency of non-holonomic control over quantum
devices with exponentially large number of Hilbert space dimensions.
We show that completely controllable devices of this type can be assembled from
elementary units of arbitrary physical nature,
and can be employed efficiently for universal quantum computations
and simulation of quantum field dynamics. 
As an example
we describe a toy device that can perform Toffoli-gate transformations
and discrete Fourier transform on 9 qubits.
\end{abstract}

\begin{multicols}{2}
\narrowtext
\section{Introduction}

What is the difference between a classical and a quantum device?
Clearly it is not in the physical laws governing their dynamics,
since Classical Mechanics follows from Quantum Mechanics as a limiting case,
when mechanical  action for each degree of freedom is much larger than the
Planck constant $\hbar$.
Hence, all classical devices are quantum as well, and the basic difference
between them is rather in the quantities of interest and in the interactions
under control.
Typically, the operators of main physical quantities have smooth dependence of
their semiclassical matrix elements on the indices numerating the energy
eigenstates, and therefore a state of the device is characterized by the
position of the center of the Ehrenfest wave packet in phase space.
The average quantities are determined as functions of this position,
whereas the finite packet width results in  uncontrolled ``quantum noise" and
is considered as an obstacle for the correct operation of the classical device
in the quantum limit. 

The situation is different in the essentially quantum limit,
where the action for each degree of freedom is of the order of $\hbar$.
Then, the matrix elements are not smooth anymore,
and the consistent description of an $N$-level device relies not only on 
quantum averages of operators, but also on all their higher moments as well.
Such description requires exhaustive information about the state of the system,
as given by a vector in the $N$-dimensional Hilbert space of the system.
Building a completely controlled quantum device in practice
implies control over all the moments and therefore is a challenging task.
It promises, however, adequately important practical benefits:
coherent control of molecules, quantum cryptography, and quantum computation
are some of the potential applications
\cite{Feynman:82,Huang:83,Deutsch:85,Tannor:85,Shapiro:86,Peirce:88,Vogel:93,Lloyd:93,DiVincenzo:95,Cirac:95,Harel:96,Law:96,Ekert:96,Harel:99,Gershkovich:00}.

In this paper we describe a scheme for constructing completely controllable
quantum devices.
We show that quantum systems perturbed in a certain time-dependent
way become ``non-holonomic",
which means that as a result of the perturbation all global constraints on the
dynamics are removed and the system becomes fully controlled
(Sec.~\ref{nhc}).
We then describe a simple, completely controllable ``unit cell''
that can serve as a building block for compound devices of arbitrary
size (Sec.~\ref{cell}),
and show in particular that it can implement the Toffoli gate
(Appendix \ref{appendix-a}).
We give examples of compound devices that can be employed efficiently for
universal quantum computations and simulation of quantum field dynamics
(Sec.~\ref{devices}).
Finally,
we describe a toy device that can perform quantum computations on 9 qubits
and show in particular how it can
perform the discrete Fourier transform on 9 qubits (Sec.~\ref{toy}).

\section{Non-holonomic Control}
\label{nhc}

The idea of controlling a system by forcing it to have
globally unconstrained---non-holonomic---dynamics 
is natural, since in order to ensure an arbitrary evolution one has first to
get rid of the restrictions posed by the existing integrals of motion and all
other constraints.
In the non-holonomic control scheme,
the system evolution is determined by an unperturbed Hamiltonian $\hat H_0$
and a number of perturbations $C_i\hat P_i$ of fixed operator structure 
$\hat P_i$ and controllable strengths $C_i$
that are applied to the system,
so that the evolution is given by the time-dependent Hamiltonian 
\begin{equation}
\hat H(t)=\hat H_0+\sum_i C_i(t)\hat P_i .
\label{Ht}\end{equation}
The system becomes non-holonomic and completely controllable if
the commutators of all orders of $\hat H_0$ and the $\hat P_i$
span the space of Hermitian operators in the Hilbert space of the system,
that is, if an arbitrary Hermitian operator can be represented as a
linear combination of the operators
\begin{equation}
\hat H_0, \hat P_i, [\hat H_0,\hat P_i], [\hat P_i,\hat P_j],
 \dots, [\hat P_i,[\hat P_j,\hat P_k]],\dots.
\end{equation}
Note that at most $N^2$ linearly independent terms are needed
for an $N$ dimensional Hilbert space.

The control scheme consists of two steps:
(i) verification that the perturbations induce non-holonomic dynamics;
and (ii) finding particular time dependencies for
the perturbations that effect a given desired control. 
Step (i) is straightforward---by inspecting the
commutation relations between the explicitly written
Hamiltonian and perturbation operators,
one checks if the system under consideration is indeed non-holonomic.
But step (ii) requires more art---one has to put the system in such
conditions that all unwanted outcomes, present in abundance in a system with
no constraints, experience a destructive quantum interference.

Let us consider a quantum system of $N=2^n$ levels,
composed of $n$ interacting two-level subsystems.
To be specific we speak about two-level atoms in a laser field, although
it could as well be any other quantum object, such as interacting spins in a
magnetic field, Josephson junctions, Rydberg atoms, rotating molecules,
quantum dots on a surface, etc.
The only requirement is that the object must be subjected to a non-holonomic
control, since only in this case it can perform any desired operation,
no matter what the physical interactions in the system are.
The choice of a practical realization of a non-holonomic system
will therefore depend mainly on optimization of technical parameters
such as simplicity and cost-effectiveness of construction,
lifetime of quantum coherence \cite{footnote decoherence},
precision of available controls, and so on.

A crucial issue that determines the strategy of construction is the required
extent of immediate universality of the control.
In principle, one can think about complete and direct physical control over a
$2^n$-level quantum system, even for a large $n$, which implies the ability to
ensure an arbitrary evolution of the system, given by any predetermined
$2^n\times 2^n$ unitary matrix $\hat U$,
and which requires $4^n$ physical control parameters.
For this purpose one should find an algorithm that determines
these controls for any given $\hat U$.
It might be difficult to find such algorithm, and even if found, its application
will require an enormous computational work that grows exponentially with
$n$, and will therefore be intractable.
In addition, the cost of physically implementing the huge number of $4^n$
control parameters seems too high a price to pay for this kind of universality,
which may not even be needed for practical purposes.
For these reasons, one should presumably give up direct universality and search
for specialized ways to build quantum devices for each particular task,
with number of controls that is not exponentially larger then what is
specifically needed. 

\vspace{-.3cm}
\section{Completely Controlled Unit Cell}
\label{cell}
\vspace{-.2cm}

One way to construct a completely controlled but not immediately universal
quantum device is to build it up from small parts, ``unit cells", each of
which is non-holonomic and therefore directly and universally controllable.
The proper functioning of the device relies then on the appropriate
connection of the cells \cite{footnote1}.
In this way the universality of the device is obtained indirectly,
not by applying a huge number of controls,
but by smartly connecting the cells and choosing 
the sequence of operations performed.
There is no general prescription how to construct a particular device;
this requires expertise in the art of ``programming" the operations
of the cells and their interactions.

\vspace{-.3cm}
\subsection{Cell structure}
\vspace{-.2cm}

\begin{figure} 
\centerline{ \psfig{file=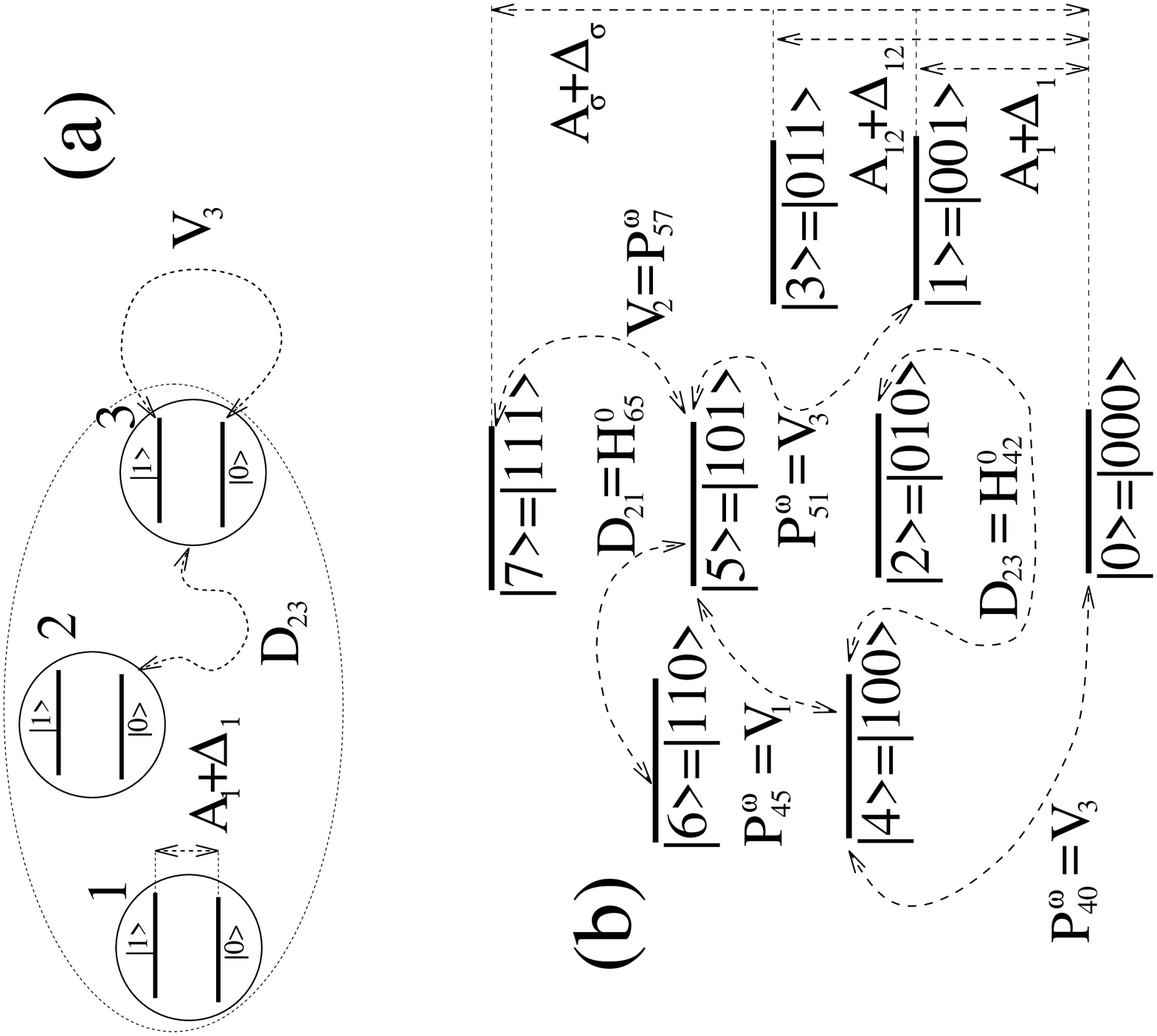,width=3.0in,angle=-90} }
\vspace{+.8cm}
\caption{
Realization of a unit cell:
A compound system of three two-level atoms interacting with external
electromagnetic and static electric fields.
(a) The $i$th atom has ground and excited states $|0\rangle_i$ and $|1\rangle_i$
with excitation energy $A_i+\Delta_i$ that can be modified by the static field;
transition amplitude in the electromagnetic field is $V_i$;
the dipole-dipole coupling of the $i$th and $j$th atoms is $D_{ij}$.
(b) The computational basis states and their relation to matrix elements of the
principal Hamiltonian $\hat H_0$ and the perturbations $\hat P_\omega$
and $\hat P_S$ of Eqs.~(\ref{eq:one}-\ref{eq:onevs}).
\label{fig:one}}
\end{figure}

An example of a completely controlled unit cell is shown in
Fig.~\ref{fig:one}.
It consists of three two-level atoms,
each with ground and excited states $|0\rangle$ and $|1\rangle$,
having distinct transition frequencies $\omega_1^a$, $\omega_2^a$,
and $\omega_3^a$.
The atoms have dipole-dipole interaction between themselves and are coupled
to two external fields: an electromagnetic field 
$E_\omega={\cal E}_\omega \cos\omega t$
of nearly resonant frequency $\omega$, and a static electric field
$E_S$.
The dipole-dipole interaction is fixed and determines the principal,
unperturbed Hamiltonian of the system, $\hat H_0$,
while the external fields provide two controllable perturbations,
$\hat P_{\omega}$ and $\hat P_{S}$.
The Hilbert space of the system has a ``computational basis" of $N=2^3=8$
states,
$|x\rangle \equiv|x_2 x_1 x_0\rangle \equiv|x_2\rangle|x_1\rangle|x_0\rangle$,
$x=0,1,\dots,7$,
where the state of the $i$th atom encodes the $i$th binary digit of
$x=\sum_{r=0}^2 x_r 2^r$ as a qubit [see Fig.~\ref{fig:one}(b)].
The crucial requirement is the non-holonomic character of the interaction.
It implies that $\hat H_0$,
$\hat P_{\omega}$, $\hat P_{S}$, and their commutators of all orders
span the linear space of $8 \times 8$ Hermitian matrices \cite{skew}.
This is indeed the case for the system shown in Fig.~\ref{fig:one},
which has principal Hamiltonian and perturbations given,
in the computational basis and assuming resonant approximation,
by the matrices 
\begin{eqnarray}
\hat H_0 &=&\left(
\begin{array}{cccccccc}
0&0&0&0&0&0&0&0\\
0&A_1&D_{12}&0&D_{13}&0&0&0\\
0&D_{21}&A_2&0&D_{23}&0&0&0\\
0&0&0&A_{12}&0&D_{23}&D_{13}&0\\
0&D_{31}&D_{32}&0&A_3&0&0&0\\
0&0&0&D_{32}&0&A_{13}&D_{12}&0\\
0&0&0&D_{31}&0&D_{21}&A_{23}&0\\
0&0&0&0&0&0&0&A_\sigma
\end{array}
\right)\!,
\label{eq:one}
\\
C_{\omega} \hat P_{\omega} &=&\left( 
\begin{array}{cccccccc}
0&V_1&V_2&0&V_3&0&0&0\\
V_1&0&0&V_2&0&V_3&0&0\\
V_2&0&0&V_1&0&0&V_3&0\\
0&V_2&V_1&0&0&0&0&V_3\\
V_3&0&0&0&0&V_1&V_2&0\\
0&V_3&0&0&V_1&0&0&V_2\\
0&0&V_3&0&V_2&0&0&V_1\\
0&0&0&V_3&0&V_2&V_1&0
\end{array}
\right)\!,
\label{eq:onevem}
\\
C_{S} \hat P_{S} &=&\left(
\begin{array}{cccccccc}
0&0&0&0&0&0&0&0\\
0&\Delta _1&0&0&0&0&0&0\\
0&0&\Delta _2&0&0&0&0&0\\
0&0&0&\Delta_{12}&0&0&0&0\\
0&0&0&0&\Delta_3&0&0&0\\
0&0&0&0&0&\Delta_{13}&0&0\\
0&0&0&0&0&0&\Delta_{23}&0\\
0&0&0&0&0&0&0&\Delta_\sigma
\end{array}
\right)\!.
\label{eq:onevs}
\end{eqnarray}
Here $D_{ij}=d_i d_j/R_{ij}^{3}$ is the dipole-dipole
coupling of the $i$th and $j$th atoms at distance $R_{ij}$, 
with $d_i$ the $i$th atom dipole matrix element,
and $V_i={\cal E}_\omega d_i$ is the dipole coupling of the $i$th atom
to the external electromagnetic field.
The excitation energy detunings of single atoms 
$A_i=\hbar (\omega_i^a-\omega)$
determine the detunings of pairs of atoms $A_{ij}=A_i+A_j$
and the total detuning $A_\sigma=A_1+A_2+A_3$.
Their values can be changed by variation of a static electric field
${E_S}$ (Stark effect), which results in energy shifts $A_i\rightarrow A_i+\Delta_i$ for single atoms,
where $\Delta_i=\alpha_i E_S$ depend on atom-specific electric 
permeability constants $\alpha_i$,
and similar shifts $\Delta_{ij}=\Delta_i+\Delta_j$ and
$\Delta_\sigma=\Delta_1+\Delta_2+\Delta_3$ for two and three atomic 
detunings respectively.

Note that by a proper choice of $\Delta _i$ and $\omega$ one can set two of
the three $A_i$ to zero.
Moreover, to simplify the presentation we also set to zero the third $A_i$,
which would otherwise remain just a part of $\hat H_0$.
Hence, hereafter all $\Delta_i$ denote just the deviations 
from zero resulting from the variation of the Stark field ${E_S}$.
The latter together with the amplitude ${\cal E}_\omega$ serve as time
dependent control parameters, $C_S$ and $C_\omega$ respectively.
The matrices $\hat P_S$ and $\hat P_\omega$ contain therefore only the
permeabilities $\alpha_i$ and the dipole moments $d_i$ respectively. 
\vspace{+.5cm}

\subsection{Cell control}

To exert direct universal control over the unit cell we proceed as follows.
(i) We fix $N^2=64$ consecutive time intervals of equal duration $T$ in which
the two perturbations will be applied to the system in an alternating sequence:
in the $k$th interval the perturbation is
$\hat P_k=\hat P_S$ for odd $k$ and $\hat P_k=\hat P_\omega$ for even
$k$, where $k=1,2,\dots,64$.
The strength of $\hat P_k$ is a controllable parameter,
which we take to have a constant value $C_k$, and which denotes either 
${\cal E}_{\omega}$
or $E_S$, during the $k$-th time interval, depending on the parity of $k$ \cite{CiPiCkPk}.
Thus, the system evolution is given by a Hamiltonian which is constant
in each interval:
\begin{equation}
\hat H(t)=\hat H_0+C_k\hat P_k\ \ \ \ \ \ t\in[(k-1)T,kT].
\end{equation}
(ii) We find $64$ positive $C_k$ values for which the total evolution of the
system will be the identity transformation: 
\begin{equation}
\hat U(t=64T)\equiv\prod_{k=1}^{64}
 \exp\left[-\frac{i}{\hbar}(\hat H_0+C_k\hat P_k)T\right]=\hat I.
\label{eq:unity}\end{equation}
To this end, we first solve the ``$8$th root" of Eq.~(\ref{eq:unity}),
\begin{equation}
\hat U(t=8T)\equiv\prod_{k=1}^{8}
 \exp\left[-\frac{i}{\hbar}(\hat H_0+C_k\hat P_k)T\right]=\hat I^{1/8},
\label{eq:nrootunity}\end{equation}
by minimizing the coefficients of the characteristic polynomial of
$\hat U(t=8T)$ \cite{Harel:99}.
This gives a sequence of positive values, $C_1,C_2,\dots,C_8$, 
for which $\hat U(t=8T)$ has the eigenvalues $e^{2\pi i m/8}$, $m=1,2,\dots,8$,
and hence satisfies $[\hat U(t=8T)]^8=\hat I$ nondegenerately.
Repeating this sequence 8 times we obtain the required 64 $C_k$.
(iii) Now,
by small variations $\delta C_k$ of the $C_k$ values we can obtain any
unitary transformation $\hat U_\epsilon$ in a small neighborhood of the
identity transformation:
\begin{equation}
\hat U(t=64T)\equiv\prod_{k=1}^{64}
 \exp\!\left[-\frac{i}{\hbar}(\hat H_0+[C_k+\delta C_k]\hat P_k)T\right]
 \!=\hat U_\epsilon.
\label{cdlup}\end{equation}
Indeed, we can present this ``small'' transformation as
\begin{equation}
\hat U_\epsilon=\exp(-i\hat{\cal H}\epsilon),
\label{uepsilon}\end{equation}
with dimensionless $8\times 8$ Hermitian Hamiltonian
$\hat{\cal H}$ which is bounded as $||\hat{\cal H}||\leq1$
and is multiplied by a small parameter $\epsilon>0$.
Now the variations $\delta C_k$ are determined to first order in $\epsilon$ by
the linear equations
\begin{equation}
\sum_{k=1}^{64}\frac{\partial \hat U(t=64T)}{\partial C_k}\,\delta C_k
 = -i\hat{\cal H}\epsilon.
\end{equation}
Moreover, when $\epsilon$ is sufficiently small,
iterative Newton method refinements of the $\delta C_k$ yield
$\hat U(t=64T)=\hat U_\epsilon$ with utmost accuracy \cite{Harel:99}.
(iv) Finally, to perform an arbitrary unitary transformation
we again present it as
$U_\epsilon$ in Eq.~(\ref{uepsilon}), but now the parameter $\epsilon$ may take
any value in $[0,2\pi]$ and will not necessarily be small.
We effect the $\hat U_\epsilon$ by dividing it into ``small'' steps:
we apply the transformation $\hat U(t=64T)=\hat U_{\epsilon/m}$ repeatedly
$m$ times, with $m$ big ($\epsilon/m$ small) enough to allow direct control,
and obtain
\begin{equation}
\hat U(t=64mT)=(\hat U_{\epsilon/m})^m=\hat U_\epsilon. 
\end{equation}
We note that, for the problem under consideration,
direct control is typically attainable with $m\leq 16$.
Moreover, we expect that $m=1$ will be sufficient
with more powerful numerical methods for solving Eq.~(\ref{cdlup})
\cite{Gershkovich:00}.
Thus, in contrast with earlier control schemes
\cite{Deutsch:85,Lloyd:93,DiVincenzo:95},
the desired unitary transformation is effected within a few control cycles,
with accuracy that depends in principle only on the physical
precision of the controls.

\begin{figure} 
\centerline{ \psfig{file=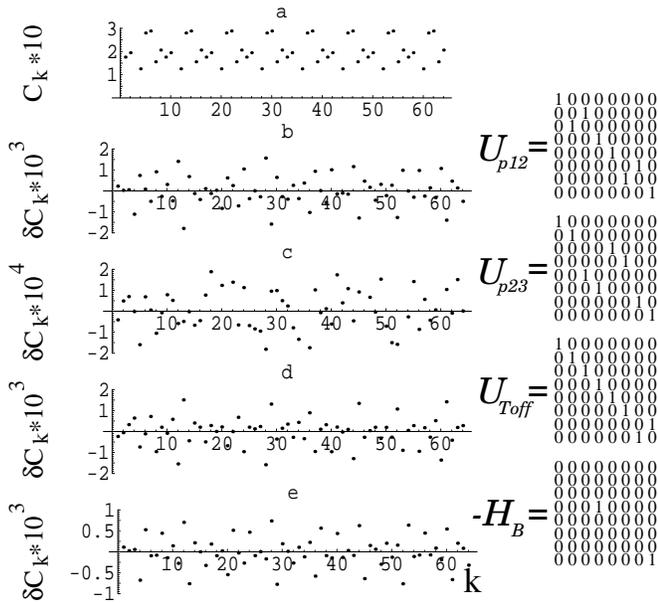,width=3.5in,angle=-90} }
\vspace{+.2cm}
\caption{
(a) Control parameters $C_k$ for the identity transformation $\hat I$.
Variations $\delta C_k$ effecting on the cell the transformation
$\hat U_{\epsilon/8}$,
with $\hat U_{\epsilon}\equiv(\hat U_{\epsilon/8})^8$ equal to:
(b) the permutation $\hat U_{p12}$;
(c) the permutation $\hat U_{p23}$;
(d) the Toffoli-gate transformation $\hat U_{Toff}$.
(e) Variations $\delta C_k$ effecting the conditional phase shift 
$\hat B(\phi)=\exp(-i\phi\hat{\cal H}_{B})$, at $\phi=\pi /32$,
employed in the quantum discrete Fourier transform.
\label{fig:two}}
\end{figure}

In Fig.~\ref{fig:two} we show examples of unit cell control,
where appropriately chosen parameters $C_k$ and variations $\delta C_k$ effect
unitary transformations on the unit cell:
the Toffoli-gate transformation (see Appendix~\ref{appendix-a}),
two-qubit permutations
$\hat p_{ij}|a\rangle_i|b\rangle_j=|b\rangle_i|a\rangle_j$ $(a,b=0,1)$,
and the conditional phase shift
employed in the quantum discrete Fourier transform
(discussed in Sec.~\ref{toy}).
The transformation is achieved either directly $(m=1)$
or by 8 repetitions $(m=8)$.
The operators $\hat H_0$, $\hat P_\omega$ and $\hat P_S$ are chosen with
arbitrary realistic values.
We take $D_{12}=1.1 E_u$, $D_{23}=0.946 E_u$, $D_{13}=0.86 E_u$,
and $T=250 \hbar /E_u$, where $E_u\sim 10^{-18}$ erg is the typical energy
scale.
For odd $k$ we switch off the external electromagnetic field, $V_{1;2;3}=0$,
and tune the atomic excitation energies by the Stark field
$E_S$ such that $\Delta_{1;2;3}=(0.1;0.11;0.312) E_u$.
For even $k$ we set $E_S=0$, that is $\Delta_{1;2;3}=0$,
and take $V_{1;2;3}=(0.3;0.33;0.24) E_u$.

\section{Completely Controlled Quantum Devices}
\label{devices}

\begin{figure} 
\centerline{ \psfig{file=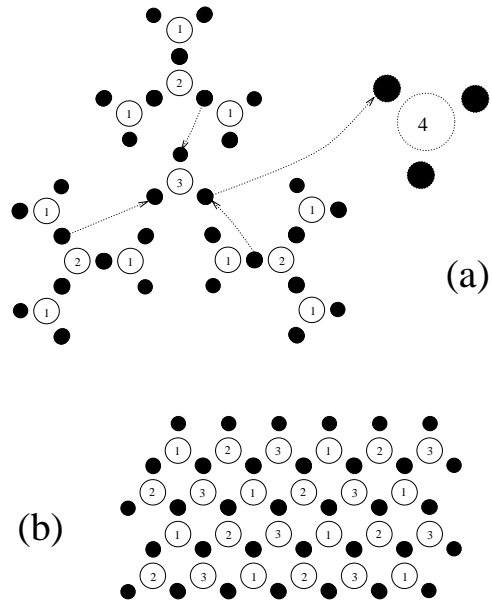,width=2.5in,angle=-90} }
\vspace{+.2cm}
\caption{
Two possible arrangements of cells for special purpose devices:
(a) tree-like structure for quantum computation;
(b) planar lattice for simulating dynamics of quantum fields.
The circled numbers denote the rank of joints of the tree (a)
or specify the order in which atoms are grouped into triads (b).
The arrows show state exchange to parent joints.
\label{fig:three}}
\end{figure}

Once completely controlled unit cells can be constructed,
a compound device can be assembled from such elements.
To be efficient, the architecture of the device will depend on the specific
function it should perform.
In Fig.~\ref{fig:three} we show two possible arrangements of unit cells for
special purpose devices:
the first arrangement suites more the purpose of quantum computing,
while the second is more useful for simulating lattice quantum field dynamics. 

The first device (Fig.~\ref{fig:three}(a)) is organized in a tree-like
structure, where the quantum state of one atom in each cell can be
exchanged with the state of an atom at the closest parent
joint of the tree.
The simplest way to make the exchange is to displace the atom to the parent
joint, however, the exchange or transport of the state without moving the
atom can be more practical.
The tree-like architecture and the possibility to perform all the unitary
transformations, including all the permutations, in each unit cell allow one
to put together and make interfering the states of any three two-level atoms of the device after at most $s=6\log_3 n$ state exchanges,
by moving them toward the root of the tree to a common cell.
Placing the new states back (if needed) requires the same number of inverse
exchanges.
This is a very modest number, $s\sim 40$, even for a rather large device
of $n\sim 10^3$ with Hilbert space of $N=2^n\sim 10^{300}$ dimensions.
Hence, all basic operations of quantum computation can be performed on any
physical system comprised of non-holonomic triads of two-level subsystems
in a tree-like structure,
and each operation can be completed within $64\times 16\times 12\times\log_3 n$
control intervals $T$.
Note that the unity transformation should be applied to all other cells
to preserve their states during the operation.


The second arrangement of cells (Fig.~\ref{fig:three}(b))
is intended mainly for emulating the dynamics of quantum fields on lattices.
Of course, it can also perform general operations on any triad,
but for a higher cost of $s=O(n^{1/2})$.
In this arrangement, after each control period of $64 T$ the closest
neighboring atoms are differently regrouped in triads (cells),
with the original grouping repeating itself after three consecutive periods.
Therefore, at each moment the change of the cell state depends on the states
of the neighboring cells, as it should be in order to emulate the dynamics of
the fields.
Immediate analogy to the Ising model emerges when we restrict ourselves to 
small values of $\epsilon$ where terms of order $\epsilon^2$ are negligible,
and then each $64T$ period plays the role of the time increment
$\Delta \tau=\epsilon$.
The evolution of such device is determined by three sums of effective
cell Hamiltonians, $\hat H_{eff}^{(p)}= \sum_{q} \hat{\cal H}_{q,p}$,
one for each period $p=1,2,3$, where $\hat{\cal H}_{q,p}$ is the effective
Hamiltonian of the $q$th cell at the $p$th period.

We can cast the cell Hamiltonians to sums of tensor products of Pauli matrices
$\hat\sigma^i_\alpha$, where the Greek index $\alpha=x,y,z$ denotes the
matrix type
and the Latin index $i$ specifies the two-level atom on which it acts.
Since the cells are under complete control, the coefficients of this
development can be made an arbitrary function of the time $\tau $,
and hence the effective Hamiltonian reads
\begin{eqnarray}
\hat H_{eff}(\tau ) &=& A_i^\alpha (\tau ) \hat \sigma _\alpha ^i
 +B_{(i,j)}^{\alpha \beta }(\tau ) \hat \sigma _\alpha ^i \hat \sigma _\beta ^j
\nonumber\\
 &&+\,C_{(i,j,k)}^{\alpha \beta \gamma }(\tau )
  \hat \sigma _\alpha^i \hat \sigma _\beta^j \hat\sigma _\gamma ^k ,
\end{eqnarray}
with implicit summation over repeated indices,
where $(i,j)$ and $(i,j,k)$ indicate pairs and triads of distinct atoms
that are periodically grouped in a common cell.
This Hamiltonian results in the evolution equation for the Heisenberg
operators $\hat \sigma_\alpha^i(\tau)$,
\begin{eqnarray}
\hbar \frac{d\hat\sigma_\alpha^i(\tau )}{d\tau }& = &
 {\cal A} _{\alpha ,j}^{i,\beta } (\tau ) \hat \sigma _\beta ^j(\tau )
 +{\cal B}_{\alpha ,(j,k)}^{i,\beta \gamma }(\tau )
  \hat \sigma _\beta ^j(\tau )\hat \sigma _\gamma ^k(\tau ) \nonumber\\
 &&+\, {\cal C}_{\alpha ,(j,k,l)}^{i,\beta \gamma \delta }(\tau )
  \hat\sigma_\beta^j(\tau )\hat\sigma_\gamma^k(\tau)\hat\sigma_\delta^l(\tau),
\label{eq:two}
\end{eqnarray}
where the coefficients ${\cal A,B,C}$ are determined by $A,B,C$
and the commutation relations of the Pauli matrices.
By a proper choice of the coefficients $A,B,C$
through the appropriate control sequences,
one can simulate different
linear and non-linear lattice models of quantum fields with time dependent
parameters.

\section{Toy Device}
\label{toy}

\begin{figure} 
\centerline{ \psfig{file=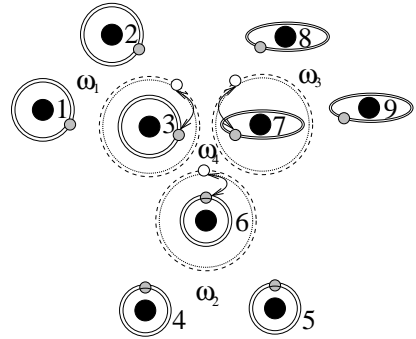,width=2.1in,angle=-90} }
\vspace{+.2cm}
\caption{
A toy device, composed of 9 Rydberg atoms,
that can perform quantum computations on 9 qubits.
Each atom is a two-level system shown schematically by double orbits.
Atoms of different triads are excited to distinct pairs of Rydberg states.
Each triad $p$ is controlled by an external field of distinct frequency
$\omega_p$.
One atom in each triad can be excited to a pair of higher Rydberg states,
thus forming a higher-level triad: (3,6,7).
These excitations (depicted by arrows) correspond to state transportations.
\label{fig:four}}
\end{figure}

We now describe a toy device that can perform quantum computations on 9 qubits.
An ensemble of 9 different Rydberg atoms is placed in a magneto-optical trap
at low temperature, as illustrated in Fig.~\ref{fig:four}.
By different atoms we mean atoms of different elements or identical atoms
that are excited to distinct pairs of Rydberg states.
The best candidates for such a device are the long-living states corresponding
to large angular momentum.
By placing all the atoms in a static electric field one lifts the degeneracy
of the magnetic quantum number and performs tuning if needed.
All the atoms experience the dipole-dipole interaction
$\hat D_{ij}=\hat d_i \hat d_j \langle R^{-3}_{ij}\rangle$,
where the cube of the inverse distance between atoms is averaged over
their translational quantum states.
Note, however, that only for almost resonant atoms this interaction
is important.
By a proper choice of the atomic states and the static field ${E_S}$,
we obtain three triads, $p=1,2,3$,
each comprised of three almost resonant two-level atoms with transition
frequencies centered around a distinct frequency $\omega_p$.
For each triad $p$, the interactions $\hat D_{ij}$ give the principal
Hamiltonian,
while a microwave field $E_{\omega_p}$ at the frequency $\omega_p$
serves as a control perturbation.
Transportation of the state of one atom in each triad to the parent joint can
be performed by dipole or Raman $\pi$ transitions from the initial pair of
Rydberg levels to a higher pair.
With these higher pairs assumed nearly resonant with a frequency $\omega_4$,
atoms 3, 6 and 7 form a higher-level triad---the parent joint of the first
three triads---which is controlled by a forth microwave field
$E_{\omega_4}$ of frequency $\omega_4$.

As an example of implementing quantum computation in the toy device,
using our non-holonomic control scheme,
we show how to perform the discrete Fourier transform modulo $N=2^9=512$
\cite{DFT}.
This is the unitary transformation on 9 qubits that is given by
\begin{equation}
\hat F_N|x\rangle=\frac{1}{\sqrt N}\sum_{y=0}^{N-1}\exp(2\pi ixy/N)|y\rangle,
\label{DFTN}\end{equation}
where $|x\rangle$ and $|y\rangle$ are states of the system computational basis.
The computational basis states are defined as
\begin{equation}
|x\rangle\equiv|x_8\rangle_9\dots|x_1\rangle_2|x_0\rangle_1,
\end{equation}
with $x\equiv\sum_{r=0}^8 x_r 2^r =0,1,\dots,N-1$ $(x_r=0,1)$,
where $|\ \rangle_i$ denotes the state of the $i$th atom---the $i$th qubit.
The algorithm we employ to perform the Fourier transform
is based on constructing the exponent in Eq.~(\ref{DFTN}) as 
\begin{equation}
\exp(2\pi ixy/2^9)=\prod_{r=0}^{8}\prod_{s=0}^{r}\exp(i\pi x_r'y_s/2^{r-s}),
\end{equation}
where $x_r'\equiv x_{8-r}$.
We begin by reversing the order in which the bits of the input $x$
are stored in our 9-qubit register,
that is, we effect the unitary transformation
\begin{equation}
|x_8\rangle_9\dots |x_1\rangle_2|x_0\rangle_1\ \rightarrow\ \  
|x_0\rangle_9\dots |x_7\rangle_2|x_8\rangle_1
\label{DFTrev}\end{equation}
by applying a sequence of state exchanges \cite{reverse}.
Then we complete the transform in 9 steps:
(i) We ``split'' the first qubit (the state of atom 1) by applying
the unitary transformation
\begin{equation}
\hat A
 \equiv\frac{1}{\sqrt2}\pmatrix{1&1\cr 1&-1}
 =\exp[\frac{-i\pi}{\sqrt8}\small\pmatrix{1-\sqrt2&1\cr 1&-1-\sqrt2}],
\end{equation}
which maps $|0\rangle\rightarrow\frac{1}{\sqrt2}(|0\rangle+|1\rangle)$
and $|1\rangle\rightarrow\frac{1}{\sqrt2}(|0\rangle-|1\rangle)$.
Note that this would already complete the Fourier transform if we had only one
qubit.
(ii) Next, we apply to the first and second qubits the conditional phase shift
$|a\rangle_2|b\rangle_1\rightarrow e^{i\pi ab/2}|a\rangle_2|b\rangle_1$ 
$(a,b=0,1)$, given explicitly by
\begin{equation}
\hat B_{21}
 \equiv\pmatrix{1&0&0&0\cr 0&1&0&0\cr 0&0&1&0\cr 0&0&0&e^{i\pi/2}}
 =\hat B(\pi/2),
\end{equation}
where $\hat B(\phi)$ is the unitary transformation
\begin{equation}
\hat B(\phi)
 =\exp[-i\phi\pmatrix{0&0&0&0\cr 0&0&0&0\cr 0&0&0&0\cr 0&0&0&-1}].
\end{equation}
Then we ``split'' the second qubit by applying to it the transformation
$\hat A$.
This accounts for the contribution of the second most significant bit of the
input $x$.
(iii)~Similarly, in steps $i=3,4,\dots,9$
we apply the conditional phase shift
$|a\rangle_i|b\rangle_j\rightarrow e^{i\pi ab/2^{i-j}}|a\rangle_i|b\rangle_j$
$(a,b=0,1)$, that is,
\begin{equation}
\hat B_{ij}
 \equiv\pmatrix{1&0&0&0\cr 0&1&0&0\cr 0&0&1&0\cr 0&0&0&e^{i\pi/2^{i-j}}}
 =\hat B(\pi/2^{i-j}),
\end{equation}
to each pair of qubits $(i,j)$, $j=1,2,..,i-1$,
and then apply the transformation $\hat A_i\equiv\hat A$ to the $i$th qubit.
Note that after the $i$th step the first $i$ qubits store the Fourier
transform of the $i$ most significant bits of $x$.
Hence, after the 9th step the Fourier transform is completed:
\begin{equation}
\hat F_{2^9}
 =(\hat A_9\hat B_{98}\cdots\hat B_{91})\cdots
 (\hat A_3\hat B_{32}\hat B_{31})(\hat A_2\hat B_{21})(\hat A_1).
\label{DFTAB}\end{equation}
Performing these operations implies also application of state exchanges
whenever one needs to transfer the states of atoms $i$ and $j$ to 
a common unit cell for processing.
A list of control commands ($\delta C_k$ sequences)
corresponding to Eqs.~(\ref{DFTrev}) and (\ref{DFTAB})
can be written straightforwardly.
 
\section{Conclusion}
\label{conclusion}

We have shown that quantum devices with exponentially large
Hilbert space dimension can be efficiently controlled,
provided they are assembled from completely controllable unit cells
in an architecture that is optimized for the specific function they
should perform.
The unit cell can be constructed from simple quantum objects of arbitrary
physical nature: two-level atoms, nuclear spins, rotating molecules,
quantum dots, etc.
This allows to optimize critical properties such as coherence time and control
precision for practical realizations.
The only requirement is that the unit cell could be put under non-holonomic
control, i.e.,
that it could be sufficiently perturbed to have unconstrained dynamics.
This ensures that the cell can be fully controlled and made perform any desired
operation.

As a concrete example, we have considered a quantum system of $2^n$ levels,
composed of $n$ two-level atoms that are coupled by dipole-dipole interactions.
The atoms are grouped into unit cells, each consisting of three nearly
resonant atoms.
Each cell is controlled with two time-dependent perturbations:
a static electric field and an electromagnetic field nearly resonant with the
atoms.  
We have shown that any unitary transformation in the $2^3=8$ dimensional
Hilbert space of the cell can be effected within a few control cycles,
each comprising 64 applications of the perturbations with values fixed 
according to a non-holonomic control scheme.
In particular, the Toffoli-gate transformation on the cell regarded as a
3-qubit register and any permutation of the three qubits can be performed.
We have given two examples of function-specific devices that can be
assembled from such cells:
(i) By arranging the cells in a ternary tree-like structure,
we obtain a device that can perform efficient quantum
computations on $n$ qubits: any unitary transformation on any three qubits
can be effected within order of $\log_3 n$ control cycles.
We have described a toy device that can perform computations on 9 qubits,
including, for example, the discrete Fourier transform.
(ii) When the atoms are arranged in a planar lattice structure,
where at each control cycle the closest
neighboring atoms are differently grouped in triads,
we can simulate various linear and non-linear lattice models of quantum
fields with time dependent parameters.

\acknowledgments

The authors are grateful to M.~Gromov for discussions and stimulating remarks.
G.H. acknowledges support from the Foundation for Fundamental Research on Matter
(FOM), which is financially supported by the Netherlands Organization for
Scientific Research (NWO).

\appendix
\section{} \label{appendix-a}

The Toffoli-gate transformation is the unitary transformation on three qubits,
\begin{equation}
\hat U_{Toff}|x_2\rangle|x_1\rangle|x_0\rangle
 =|x_2\rangle|x_1\rangle|x_0\,{\rm XOR}\,(x_1\,{\rm AND}\,x_2)\rangle,
\end{equation}
which corresponds to the three-bit classical logic gate,
\begin{eqnarray}
x_2 &\rightarrow& x_2'=x_2 \nonumber\\
x_1 &\rightarrow& x_1'=x_1 \nonumber\\
x_0 &\rightarrow& x_0'=x_0\,{\rm XOR}\,(x_1\,{\rm AND}\,x_2),
\end{eqnarray}
introduced by Toffoli as a universal gate for classical reversible computation
\cite{Tgate}.
It acts as a permutation of the computational basis states,
$|x\rangle \equiv |x_2\rangle|x_1\rangle|x_0\rangle$,
$x\equiv\sum_{r=0}^2 x_r 2^r=0,1,\dots,7$, given by the unitary matrix
\begin{equation}
\hat U_{Toff} =\left(
\begin{array}{cccccccc}
1&0&0&0&0&0&0&0\\
0&1&0&0&0&0&0&0\\
0&0&1&0&0&0&0&0\\
0&0&0&1&0&0&0&0\\
0&0&0&0&1&0&0&0\\
0&0&0&0&0&1&0&0\\
0&0&0&0&0&0&0&1\\
0&0&0&0&0&0&1&0
\end{array}
\right).
\label{UToff}
\end{equation}
This matrix can be presented as
\begin{equation}
\hat U_{Toff}=\exp(-i\pi\hat{\cal H}_{Toff}),
\end{equation}
with the (idempotent) Hermitian matrix
\begin{equation}
\hat {\cal H}_{Toff} =\frac{1}{2}\left(
\begin{array}{cccccccc}
0&0&0&0&0&0&0&0\\
0&0&0&0&0&0&0&0\\
0&0&0&0&0&0&0&0\\
0&0&0&0&0&0&0&0\\
0&0&0&0&0&0&0&0\\
0&0&0&0&0&0&0&0\\
0&0&0&0&0&0&1&-1\\
0&0&0&0&0&0&-1&1
\end{array}
\right).
\label{HToff}
\end{equation}
In our control scheme the Toffoli-gate transformation can be effected
on the unit cell by repeating 8 times the transformation
$\hat U_{\epsilon/8}\equiv\exp(-i\pi\hat{\cal H}_{Toff}/8)$
$(\epsilon=\pi)$, which is directly attainable:
$\hat U(t=64T)=\hat U_{\epsilon/8}$ (see Fig.~\ref{fig:two}(d)).

\end{multicols}

\begin{thebibliography}{99}
\bibitem{footnote decoherence} We do not consider here the important issue of
 decoherence and error correction, since the correction procedure can be
 treated as a particular case of operating the quantum device.
\bibitem{footnote1} One can prove that a system comprised of two non-holonomic
 parts, with an interaction that cannot be represented as a single tensor
 product of operators related to each part, is in itself non-holonomic;
 see Ref.~\cite{Gershkovich:00}.
\bibitem{skew} More Precisely, $i\hat H_0$, $i\hat P_\omega$, $i\hat P_S$,
 and their commutators of all orders must span the 64-dimensional real
 linear space of $8 \times 8$ anti-Hermitian matrices.
\bibitem{CiPiCkPk} In the notation of Eq.~(\ref{Ht}), for $t\in[(k-1)T,kT]$
we have $\hat H(t)=\hat H_0+C_S(t)\hat P_S +C_\omega(t)\hat P_\omega$, with 
$C_\omega(t)=0$ and $C_S(t)=C_k$ for odd $k$, and 
$C_S(t)=0$ and $C_\omega(t)=C_k$ for even $k$.
\bibitem{DFT} The quantum discrete Fourier transform is reviewed in 
 Ekert and Jozsa, Ref.~\cite{Ekert:96}.
 An early proposal for its physical implementation was given by Cirac and
 Zoller, Ref.~\cite{Cirac:95}.
\bibitem{reverse} This reversal can be avoided if we agree to read the
 result in the bit-reversed order; see Ekert and Jozsa, Ref.~\cite{Ekert:96}.
\end{thebibliography}
\end{document}